\documentclass[%
 aip,
reprint,
 amsmath,amssymb,
]{revtex4-2}

\usepackage{graphicx}
\usepackage{dcolumn}
\usepackage{bm}
\usepackage[utf8]{inputenc}
\usepackage[T1]{fontenc}
\usepackage{mathptmx}
\usepackage{etoolbox}

\usepackage{hyperref}
\hypersetup{colorlinks=true,linkcolor=blue,filecolor=magenta,urlcolor=cyan,}

\usepackage{siunitx}
\usepackage{braket}

\usepackage{placeins}

\makeatletter
\def\@email#1#2{%
 \endgroup
 \patchcmd{\titleblock@produce}
  {\frontmatter@RRAPformat}
  {\frontmatter@RRAPformat{\produce@RRAP{*#1\href{mailto:#2}{#2}}}\frontmatter@RRAPformat}
  {}{}
}%
\makeatother

\begin{document}

\preprint{APS/123-QED}

\title{Fiber-tip endoscope for optical and microwave control}

\author{Stefan Dix}
\thanks{Both authors contributed equally; email:mdix@physik.uni-kl.de}
\affiliation{Department of Physics and State Research Center OPTIMAS, University of Kaiserslautern, Erwin-Schroedinger-Str. 46, 67663 Kaiserslautern, Germany}
\author{Jonas Gutsche}
\thanks{Both authors contributed equally; email:mdix@physik.uni-kl.de}
\affiliation{Department of Physics and State Research Center OPTIMAS, University of Kaiserslautern, Erwin-Schroedinger-Str. 46, 67663 Kaiserslautern, Germany}
\author{Erik Waller}%
\affiliation{Department of Physics and State Research Center OPTIMAS, University of Kaiserslautern, Erwin-Schroedinger-Str. 46, 67663 Kaiserslautern, Germany}
\affiliation{Fraunhofer Institute for Industrial Mathematics ITWM, Fraunhofer-Platz 1, 67663 Kaiserslautern, Germany}
\author{Georg von Freymann}%
\affiliation{Department of Physics and State Research Center OPTIMAS, University of Kaiserslautern, Erwin-Schroedinger-Str. 46, 67663 Kaiserslautern, Germany}
\affiliation{Fraunhofer Institute for Industrial Mathematics ITWM, Fraunhofer-Platz 1, 67663 Kaiserslautern, Germany}
\author{Artur Widera}

\affiliation{Department of Physics and State Research Center OPTIMAS, University of Kaiserslautern, Erwin-Schroedinger-Str. 46, 67663 Kaiserslautern, Germany}

\date{\today}

\begin{abstract}
We present a robust, fiber based endoscope with a silver direct-laser-written (DLW) structure for radio frequency (RF) emission next to the optical fiber facet. 
Thereby, we are able to excite and probe a sample, such as nitrogen vacancy (NV) centers in diamond, with RF and optical signals simultaneously and specifically measure the fluorescence of the sample fully through the fiber.
At our targeted frequency-range around $\SI{2.9}{\giga\hertz}$ the facet of the fiber-core is in the near-field of the RF-guiding silver-structure, which comes with the advantage of an optimal RF-intensity decreasing rapidly with the distance. By creating a silver structure on the cladding of the optical fiber we achieve the minimal possible distance between an optically excited and detected sample and an antenna structure without affecting the optical performance of the fiber. This allows us realizing a high RF-amplitude at the sample's position when considering an endoscope solution with integrated optical and RF access. The capabilities of the endoscope are quantified by optically detected magnetic resonance (ODMR) measurements of a NV-doped microdiamond that we probe as a practical use case. We demonstrate a magnetic sensitivity of our device of $\SI{17.8}{\nano\tesla} / \sqrt{\SI{}{\hertz}}$ when measuring the ODMR exclusively through our fiber and compare the sensitivity to a measurement using a confocal microscope. 
Moreover, such an endoscope could be used as a powerful tool for measuring a variety of fluorescent particles that can otherwise only be measured with bulky and large optical setups. Furthermore, our endoscope points toward precise distance measurements based on Rabi oscillations.
\end{abstract}

\maketitle

\section{Introduction}

\begin{figure*}
\centering
\def\svgwidth{\textwidth}
\includegraphics[width=0.9\textwidth]{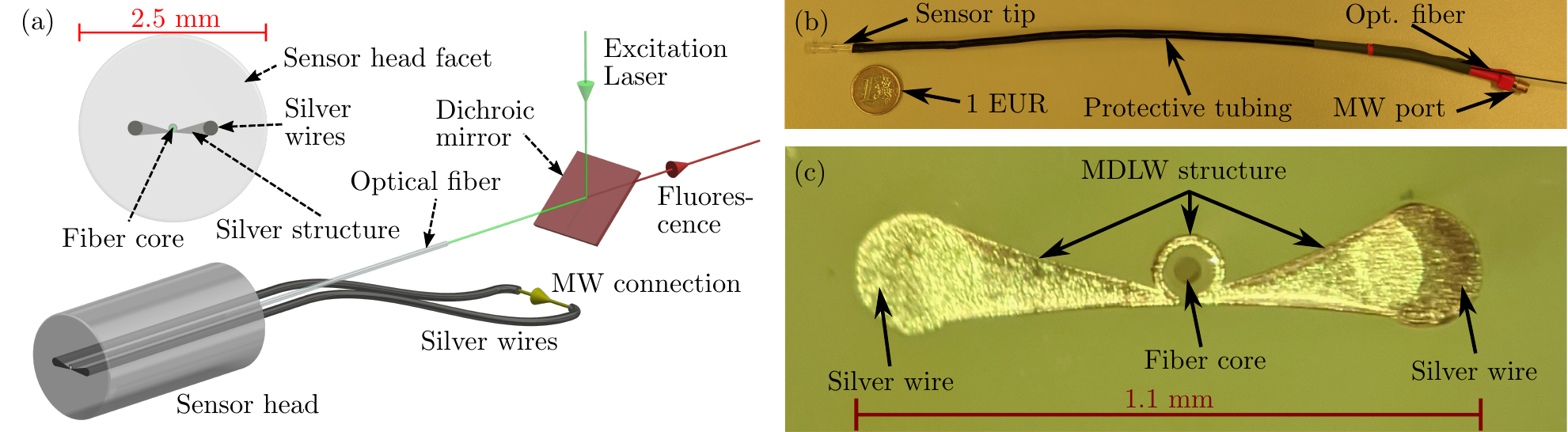}
\caption{Schematic of the fiber-tip endoscope antenna (a). A silver-based mDLW antenna is fabricated on the facet of an optical fiber and connected to additional silver wires. The MW signal is supplied to the cables via a SMA-jack which is soldered to the silver wires. Optical access for excitation and read out of NV centers is provided by the multimode fiber. An image of the whole sensor including an one euro coin as a size reference is shown in (b). In (c) a microscopic image of the final silver structure on the sensor tip is depicted.}
\label{complete_sensor}
\end{figure*}

During the last two decades the nitrogen vacancy (NV) center in diamond has been studied extensively for magnetic field sensing \cite{Rondin.2014, Boretti.2019}, thermometry in bio-medical applications \cite{Kucsko.2013}, as well as quantum information processing \cite{Liu.2018}.
As a sensor, this solid-state system can be addressed, manipulated and read out harnessing optical and microwave (MW) fields \cite{Doherty.2013} and reaches sensitivities in the order of $\eta < \SI{1}{\nano\tesla}/ \sqrt{\SI{}{\hertz}}$ \cite{Zhang.2021}. Further, it has been observed in host crystals as small as $\SI{5}{\nano\meter}$ \cite{Tisler.2009, Terada.2019} and has been introduced into living cells \cite{Kucsko.2013}.
This extreme minaturization and compatibility goes well beyond the capabilities of alkali vapor cells \cite{Budker.2007}, for example, which have been shown to exhibit even improved sensitivities of $\eta < \SI{1}{\femto\tesla}/ \sqrt{\SI{}{\hertz}}$ \cite{Kominis.2003, Keder.2014}.\newline
Usually, solid-state color centers are investigated optically in a microscope with an additional microwave setup \cite{Levine.2019}.
Therefore, a variety of fiber-tip integrations have been developed relaxing the need for bulky measurement apparatuses.
However, in most of these applications, the microwave fields, which are crucial for spin-manipulation, were supplied by an external antenna located at a comparably large distance in the order of a few millimeter to the host diamond \cite{Bai.2020,Chatzidrosos.2021,Dong.2018,Duan.2020,Fedotov.2016,Kuwahata.2020,Sturner.2021}.
In order to reduce the overall MW powers applied to an examined sample in close proximity, which are typically in the range from $\SI{}{\milli\watt}$ to several $\SI{}{\watt}$ \cite{Labanowski.2018} and to simultaneously sustain the driving field strength needed for vector magnetometry with NV centers, a coupling of the magnetic near-field component to the spins is necessary.
Here, we present the prototype of a fiber-tip antenna that allows MW manipulation as well as optical control of NV centers simultaneously.
A schematic drawing and two images of the fabricated device are shown in FIG. \ref{complete_sensor}.
It consists of a commercially available multimode fiber and two attached silver wires, which are fixed in position with epoxy resin.
The wires are connected by an $\Omega$-shaped antenna fabricated using metallic direct laser writing (mDLW) in a silver-based photo-resist on the fiber facet \cite{Waller.2021} and conductivity is increased in a galvanic post process.
Such antennas are widely applied for magnetometry with NV centers due to their homogeneous magnetic field in the center \cite{Opaluch.2021}.
The diameter of the prototype has a size of $d=\SI{2.5}{\milli\meter}$ in order to use commercially available tools for fiber processing and to fit into fiber caps for storage. 
We emphasize that the diameter can be further reduced without changing the general production steps.
The length of the prototype's fiber is $l=\SI{1.4}{\meter}$ and the unprocessed side of the fiber is connected to the optical setup used for optical access and readout.
The MW signal is fed from a microwave source to the silver wires with a standard SMA-jack.\newline
We point out that the application of the device presented here is not limited to magnetometry using NV centers. 
Harnessing the coherence properties of the NV-center, we show the working-principle of the fiber-tip antenna as a distance sensor based on the measurement of the Rabi frequency $\Omega$ in a pulsed ODMR measurement.
Moreover, the device could also be employed in the coherent microwave control of Silicon-Vacancy (SiV) centers in diamond \cite{Sukachev.2017,Zhang.2020}, Boron-Vacancy centers ($\mathrm{VB}^-$) in hexagonal Boron-Nitride (hBN) \cite{Gottscholl.2021} or Divacancies in Silicon-Carbide (SiC)\cite{Christle.2015}. 
Furthermore, the fiber-tip antenna presented here could also be a powerful tool for precise microwave control of cold gases inside \cite{Langbecker.2018} or in close proximity to an optical fiber \cite{Vetsch.2010},
as well as for excitation of spin-waves \cite{Vogel.2018}. \newline
In the following, we detail the device fabrication using mDLW on the fiber facet.
Moreover, we measure the antenna characteristics by means of scattering parameters and show that a high coupling of the magnetic field in close proximity is achieved.
Finally, we investigate the device's sensing capabilities by applying it as a NV magnetometer and distance sensor using cw and pulsed 
ODMR and compare it to a conventional microscope setup.

\section{Device Production}

\begin{figure*}
		\includegraphics[width=0.9\textwidth]{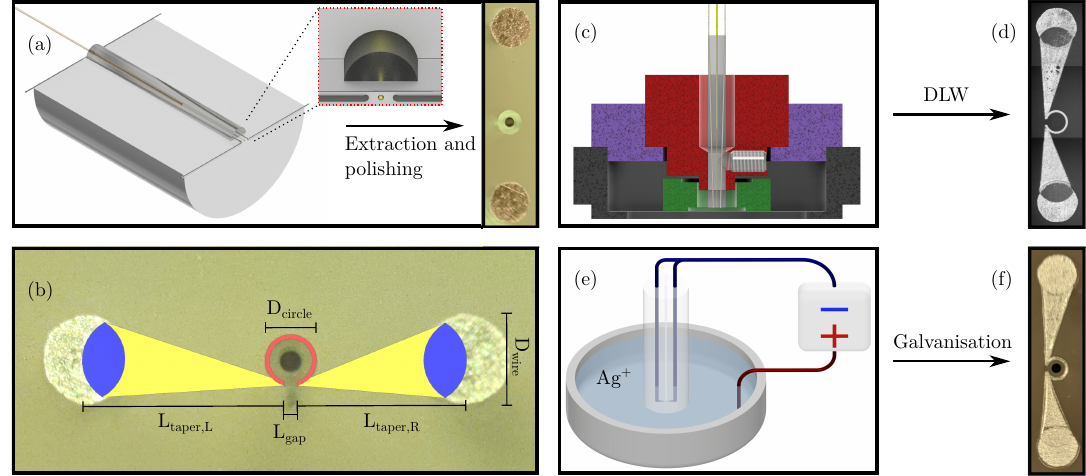}
		\caption{The production of our endoscope starts with a PTFE mold for silver wires and fiber, which is shown in the left and central image in (a). The result after extraction and polishing of the sensor tip is shown in the right section of (a). (b) shows the geometrical construction and dimensions of the silver structure drawn over an image of the fiber facet which is planned to be directly laser written on. The geometrical shape and size of the structure is adapted to the actual position of silver wires and the optical fiber core by taking images with the microscope system of the PPGT using the 20x, 0.5NA objective. For a scale reference the calibrated stage of the PPGT is utilized. The red circle has an inner diameter of $\SI{100}{\micro\meter}$ and an outer diameter of $\mathrm{D_{circle}}=\SI{120}{\micro\meter}$ with a gap distance at the bottom of $\mathrm{L_{gap}}=\SI{30}{\micro\meter}$ and is centered around the fiber core. As a connection between the circle and the outer silver wires a taper structure (yellow) is created. The blue part of the structure marks the overlapping area between the taper structure and the silver wires. The values of the relevant sizes are $\mathrm{L_{taper,L}}=\SI{469}{\micro\meter}$, $\mathrm{L_{taper,R}}=\SI{393}{\micro\meter}$ and $\mathrm{D_{wire}}=\SI{200}{\micro\meter}$. After creating the antenna structure the fiber stays mounted in the special mount (c) for the DLW process, which was also used for taking the microscope images of the sensors tip with the PPGT. The resulting structure is shown in (d). Since this is a not well enough conducting structure a galvanisation post process (e) is realised. The final structure is shown in (f).}
		\label{Steps}
\end{figure*}

In order to fabricate a compact, robust endoscope with integrated RF-antenna on an optical multimode-fiber as optical access port we chose a design for the endoscope head which is similar to a typical fiber ferrule with a diameter size of $\SI{2.5}{\milli\meter}$.
The fabrication starts with a Polytetrafluoroethylene (PTFE) mold in a cylindrical shape with $\SI{25}{\milli\meter}$ height and a diameter of $\SI{20}{\milli\meter}$ as shown as a half section schematic in FIG. \ref{Steps} (a). 
A bore hole with a drill diameter of $\SI{2.5}{\milli\meter}$ and a depth of around $\SI{22}{\milli\meter}$ is added centered in the lid surface of the cylinder.
From the other side of the cylinder three holes with a $\SI{0.3}{\milli\meter}$ drill are added, where one hole is also centered and the 2 other holes are drilled in a $\SI{0.5}{\milli\meter}$ distance left and right from the center such that the three holes form a line. 
Since PTFE is a relatively soft material the actual size of the $\SI{0.3}{\milli\meter}$ bore holes is slightly smaller and perfectly fit a $\SI{200}{\micro\meter}$ silver wire (99.99\% silver) and a $\SI{125}{\micro\meter}$ cladded fiber with a $\SI{50}{\micro\meter}$ core diameter (Thorlabs FG050LGA) without significant clearance.
The silver wires are subsequently added into the mold and taped to the sides to achieve a slightly V shaped path such that cables are not in physical contact avoiding a short circuit. 
In the next step a partially stripped fiber is added such that the end of the stripped part is roughly in the center of the mold.
The $\SI{0.3}{\milli\meter}$ holes are sealed up with a removable gum-like glue (Fixogum) and filled with epoxy resin and hardener HT2 from R\&G. 
During the 24 hours curing process at room temperature, the fiber is fastened in a centered position at the top of the mold. 
A schematic half section image of the filled mold is shown in the left part of FIG. \ref{Steps} (a).
After curing, the epoxy mold is sliced in half by sawing from both sides until the mold can carefully be broken and the epoxy with the fiber and silver cables is extracted. 
To improve the stability and avoid breaking of the fiber for the following process shrink tubing is added to increase the stiffness and ensure electrical insulation between both silver wires.
In the next step the facet of the sensor head is polished on diamond polishing paper for optical fibers in a bare ferrule fiber polishing puck (D50-F) from Thorlabs preparing a flat surface for the direct laser writing process.
The sensor's surface after polishing is shown in the right side in FIG. \ref{Steps} (a).

With the geometry given after polishing, we design an $\Omega$-shaped antenna structure consisting of a circle with a gap connected by two tapers to the silver wires, as shown in FIG. \ref{Steps} (b).
For the direct laser-writing process of such a structure, we employ a 2D silver resist consisting of silver-perchlorate, water and trisodium citrate \cite{Waller.2019}. 
A special mount for the fiber, shown in FIG. \ref{Steps} (c), is holding the fiber in position during the writing process in a small distance around $\SI{100}{\micro\meter}$ to $\SI{150}{\micro\meter}$ above a glass substrate with the silver resist on top of it. 
In order to reduce the evaporation of the silver resist, an additional ring around the fiber and the silver resist is added and sealed with Fixogum. 
The writing process is carried out on a commercial Nanoscribe system (Photonic Professional GT, PPGT) which uses a femtosecond $\SI{780}{\nano\meter}$ laser for two-photon reduction.
Using an air objective (20x 0.5NA, Nanoscribe), we write the structure in three slices with a slicing distance of $\SI{1.5}{\micro\meter}$ and hatched in $\SI{0.3}{\micro\meter}$ steps with a writing speed of $\SI{200}{\micro\meter \per {\second}}$.
The two additional slices improve the structure density, where less silver was deposited due to surface roughness of the fiber tip during the writing process or microscopic holes and cracks in the surface.
The final silver structure consists of elementary silver and is developed by rinsing water over the sensor head and blow-drying with compressed air. 
The DLW structure connecting both silver wires and forming the antenna on the fiber facet is shown in FIG. \ref{Steps} (d).
The thickness of the DLW structure is approximately in the order of below $\SI{1}{\micro\meter}$, because the 2D resist cannot provide higher structures, and a 2.5D resist failed in producing sufficiently high quality structures due to overheating.
Electrical resistance measurements after the development process of the structure show an ohmic resistance in the $\SI{}{\kilo\ohm}$ to $\SI{}{\mega\ohm}$ range, which can be explained by the thin silver layer in combination of microscopic cracks and holes at borders between different materials and in the surface of the fiber tip. 
To improve the conductivity of the structure, an additional galvanisation step is performed as depicted in FIG. \ref{Steps} (e).
To this end, a commercial galvanisation solution from Marawe is used. 
The solution consists of a silver-electrolyte and a surfactant for a smooth surface.
The voltage in the galvanic solution is set to $\SI{0.8}{\volt}$ and the structure is kept in the solution for 6 minutes in total and then rinsed with water.
A silver-wire ($\SI{99.99}{\percent}$) with a diameter of $\SI{200}{\micro\meter}$ serves as a cathode. 
After galvanisation the final antenna structure, which is shown in FIG. \ref{Steps} (f), has an ohmic resistance of $R=\SI{0.51}{\ohm}$ measured through both silver wires and the structure itself with a digital multimeter (Keithley 2100 6 1/2 DIGIT MULTIMETER) after subtraction of the resistance of the connecting cables.
The thickness of the silver structure $z$ is subsequently measured with an optical microscope setup described below using an objective with a NA of $0.5$.  
The distance $z$ between the focus points of sharp images of the antenna surface and the fiber facet surface is used to estimate the thickness. Hereby, a thickness of the structure of about $z \approx \SI{5}{\micro\meter}$ is determined.\newline
For the optimization of the antenna's resonance frequency the length of the silver wires is cut to a multiple of the wavelength $\lambda_\mathrm{res}$ at the resonance frequency of $f_\mathrm{res} = \SI{2.87}{\giga\hertz}$ which corresponds to $\lambda_\mathrm{res} \approx \SI{104.46}{\milli\meter}$. 
Subsequently, the wires are soldered to a SMA-jack and the resonances of the antenna are measured with a vector network analyzer (VNA) MS2038C from Anritsu. 
The resonance frequency of the antenna is adjusted by modifying the length of the silver cables and measuring the frequency response using the VNA.
Further shrink tubing and furcation tubing is then added to the whole cable length to reduce the flexibility of the device and increase the stability of all components and the resonance frequency of the antenna which can shift due to external stress on the sensor or bending of the wires. 
As a last step, a FC/PC adapter is connected to the end of the optical fiber enabling a quick connection of the sensor to a laser source via a fiber coupler.

\section{Antenna Characteristics} \label{sec:characterization}

\begin{figure}
	    \centering
		\includegraphics[width=0.45\textwidth]{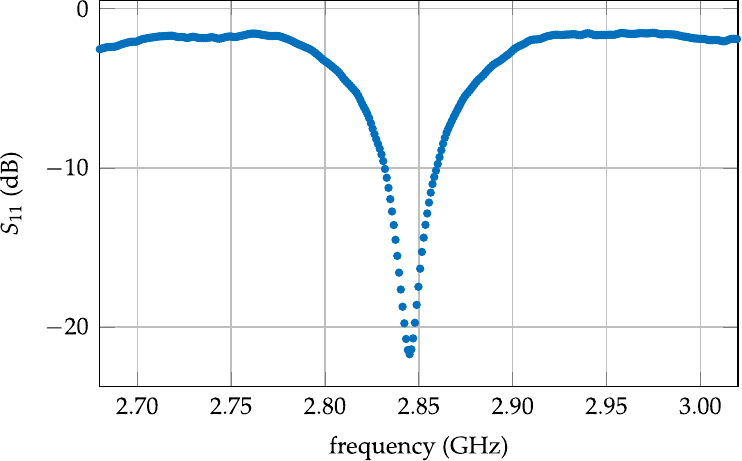}
		\caption{Experimental scattering parameter $S_{11}$ in dependence of the frequency. The minimum occurs close to the aimed frequency to drive the NV centers spin transitions at $\SI{2.87}{\giga\hertz}$.}
		\label{fig:antennachar}
\end{figure}

As a first step for the characterization of the produced antenna-structure the scattering parameter $S_{11}$, a value for the reflected power of the antenna, is measured with the VNA. 
For $S_{11}$, a small value is desirable corresponding to a large fraction of the power irradiated from the antenna. 
For a two-port measurement, the scattering parameters determined by the VNA are defined as 
\begin{equation}
\begin{pmatrix}
b_1\\
b_2
\end{pmatrix}
=
\begin{pmatrix}
S_{11} & S_{21}\\
S_{12} & S_{22}
\end{pmatrix}
\begin{pmatrix}
a_1\\
a_2
\end{pmatrix}
,
\end{equation}
where $a$ represents the incoming wave at each antenna, $b$ the reflected wave and the index is the number of the port of the respective antenna. 
The reflection parameter $S_{11}$ in the area of the expected resonance of an NV center around $\SI{2.87}{\giga\hertz}$ is shown in FIG. \ref{fig:antennachar}.
We observe a reduction of the $S_{11}$ parameter to $\SI{-22}{dB}$ within a bandwidth ($\SI{-3}{dB}$) of $\Delta \nu_\mathrm{1,\SI{-3}{dB}}=\SI{6.9}{\mega\hertz}$ around a resonance frequency of $\nu_\mathrm{1,res}=\SI{2.846}{\giga\hertz}$ of the antenna. 
Further, we observe shifts in the order of few $\SI{10}{\mega\hertz}$ in the antenna's resonance frequency due to bending of the RF guiding silver wires and due to the presence of a close by detector for example a second antenna in close proximity, which is shown in FIG. \ref{fig:antennachar_app} in appendix \ref{app:s_parameter}.
In the proposed example application of magnetometry with NV centers, the Zeeman-shift due to a magnetic bias-field is typically several orders of magnitude larger than the bandwidth of the antenna \cite{Doherty.2013}. 
Therefore, in broadband applications as ODMR-measurements with NV centers, antennas are often driven off resonance and we use the high field intensity in close proximity to the silver DLW structure to efficiently couple to the magnetic dipole transitions of the spin-states. 
In addition to further measurements in chapter \ref{device_prop}, which determine the microwave's magnetic field intensity as a function of distance to the antenna structure, a measurement of $S_{21}$ is shown in FIG. \ref{fig:antennachar_app} in appendix \ref{app:s_parameter}. 
Since measurements with a VNA, such as the ones shown additionally in FIG. \ref{fig:antennachar_app}, cannot fully distinguish between the electrical or magnetic field components and macroscopic detectors cannot be considered point sources, we utilize diamonds containing NV centers and Rabi oscillations as a detection method to characterize the distance dependency of the magnetic field intensity emitted by the antenna, which is described in detail in chapter \ref{device_prop}. 
In the course of these measurements, we have verified that the antenna properties are unchanged, if powers of up to $\SI{40}{dBm}$ are applied in measurements taken over a period of several weeks in total.

\section{Test setup for measurements with NV centers}

\begin{figure*}
    \centering
    \includegraphics[width=.9\textwidth]{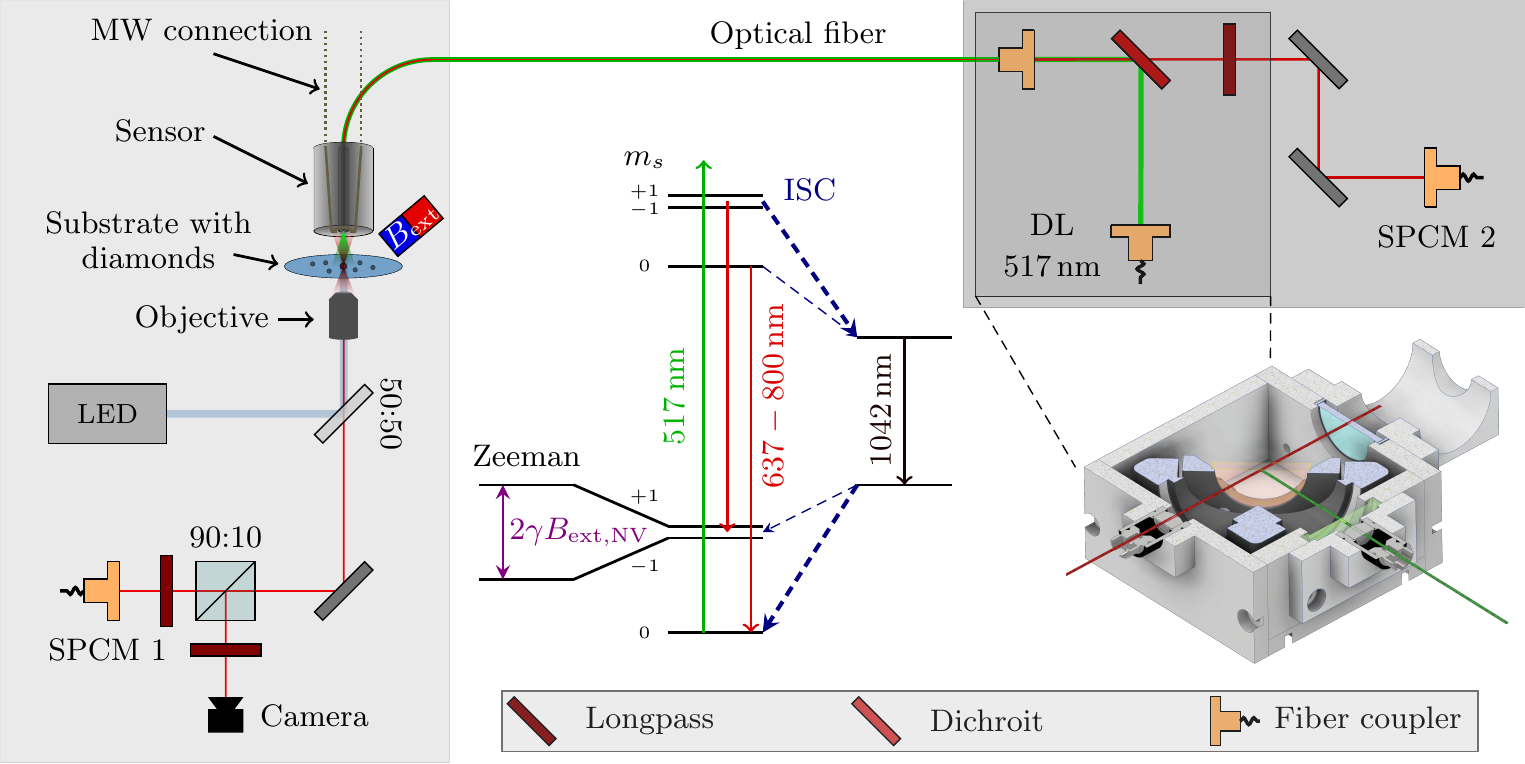}
    \caption{Overview of the measurement setup including the level scheme of the NV$^-$ center. A diode laser (DL) is coupled into the optical fiber of the endoscope and microwaves are applied to the mDLW-antenna at the MW connection wires. The fluorescence light of NV-centers in a single-crystalline microdiamond is either gathered with the microscope objective into the confocal beam path or with the endoscope. In both paths, the fluorescence light is transmitted to a fiber-coupled single-photon counting module (SPCM). This setup enables simultaneous detection in both beam-paths and, therefore, a comparison between confocal microscope and endoscope.}
    \label{fig:Setup_schematic}
\end{figure*}

In order to test the device produced, we employ additionally a microscope for NV magnetometry as depicted in FIG. \ref{fig:Setup_schematic}.
The sample examined consists of a glass substrate with $\SI{15}{\micro\meter}$-sized microdiamonds hosting NV centers with a doping of $c_{\mathrm{NV}} = \SI{3.5}{ppm}$ from Adamas Nanotechnologies. 
The NV center is a paramagnetic, optically active point defect in the diamond lattice composed of a substitutional nitrogen atom (N) and an adjacent vacant lattice site (V) \cite{Doherty.2013}.
This defect features local energy levels in the diamond's band-gap, which can be addressed with optical and microwave wavelengths, as depicted in the level-scheme in FIG. \ref{fig:Setup_schematic}. 
Its ground state splits into a spin-triplet with $D = \SI{2.87}{\giga\hertz}$ zero-field splitting between the $m_s=0$ and degenerate $m_s=\pm 1$ spin states, quantized along the NV-axis. This $m_s=\pm1$ degeneracy is split by crystal strain $E$ and the presence of an external magnetic field $\vec{B}_{\mathrm{ext}}$. Ground-state splitting and Zeeman shift, caused by the magnetic field component along the NV-axis, can be described by the Hamiltonian \cite{Doherty.2013,Rondin.2014}
\begin{equation}
    H=h D S^2_z + hE(S_x^2-S_y^2) + \gamma \vec{S} \cdot \vec{B}_{\mathrm{ext}},
\end{equation}
where $h$ is the Planck constant, $\vec{S}=(S_x,S_y,S_z)^T$ is the vector of the Pauli matrices and $\gamma = \mu_B g_s / h$ is the splitting constant with the Bohr magneton $\mu_B$ and Lande-factor $g_s$.
The optically excited state is energetically separated by $\SI{1.945}{e\volt}$ corresponding to a zero phonon line (ZPL) at $\lambda_{\mathrm{ZPL}}=\SI{637}{\nano\meter}$.
The line is strongly broadened by phononic states of higher energy allowing for off-resonant excitation at room temperature.
Thus, the excited states can decay under emission of a photon in the phonon-broadened band ranging from $\SI{637}{\nano\meter}$ to $\SI{800}{\nano\meter}$. 
Further, a second decay path via intersystem-crossing (ISC) into long-lived singlet states and a subsequent second ISC back to the triplet ground state exists, leading to infrared emission of $\SI{1042}{\nano\meter}$ outside the measured fluorescence band. The $m_s=\pm1$-states preferably decay via this path, and, therefore, optical pumping leads to spin-polarization into the $m_s=0$ ground state.
The same mechanism is harnessed for optical magnetic resonance (ODMR) of the spin states under resonant microwave excitation. 
Here, the spin-polarized NV centers are subject to a coherent manipulation of the ground state via microwaves in the range of $\SI{2.87}{\giga\hertz}$ reducing the optical fluorescence dependent on the spin state population \cite{Doherty.2013,Rondin.2014}.

The fabricated fiber-tip endoscope is precisely positioned in close proximity of a single-crystalline microdiamond using the camera of the wide-field microscope.
A microwave setup for cw and pulsed excitation is connected with the SMA-jack of the antenna produced.
Moreover, a laser source (DLnsec 520, LABS electronics) with a wavelength of $\lambda=\SI{517}{\nano\meter}$, which can be operated in cw and pulsed mode is fiber-coupled to the connection box and coupled into the endoscope with a dichroic mirror (DMLP 605, Thorlabs).
Therefore, the fiber-tip endoscope can drive optical as well as microwave transitions of NV centers.
Further, the optical response in ODMR measurements, i.e., the intensity of the fluorescence light, is collected via the fiber-endoscope as well as with the confocal microscope configuration. 
In the confocal beam-path, we employ a longpass-filter (FELH650, Thorlabs) to transmit NV-fluorescence to a fiber-coupled single-photon counting module (SPCM, Count-T-100, lasercomponents) for detection. 
The light collected by the fiber-tip endoscope is guided back into the connection box and NV fluorescence is transmitted through the dichroic mirror and additionally filtered by a longpass-filter (FELH 650).
Furthermore, the fluorescence light is measured either directly with a photodiode, or by applying a fiber-coupled sensor, which is a second SPCM (Count-T-100, lasercomponents) in our setup.
All optics connected to the fiber-tip endoscope for excitation and read-out of NV centers can be mounted in the small connection box and accessed via multiple optical fibers.
This leads to a portable device enabling quick exchange of individual filters, excitation lasers and detectors.
The whole configuration allows simultaneous detection and comparison of the ODMR-signal gathered with the fiber and the confocal microscope.

\section{Device properties} \label{device_prop}
\subsection{CW ODMR}
We apply the fiber-tip endoscope to record a spectrum of the electron spin resonances of NV centers in an arbitrary external magnetic field via continuous-wave ODMR, using a cw laser source and sweeping the microwave frequency across the resonances.

\begin{figure*}
	    \centering
		\includegraphics[width=0.9\textwidth]{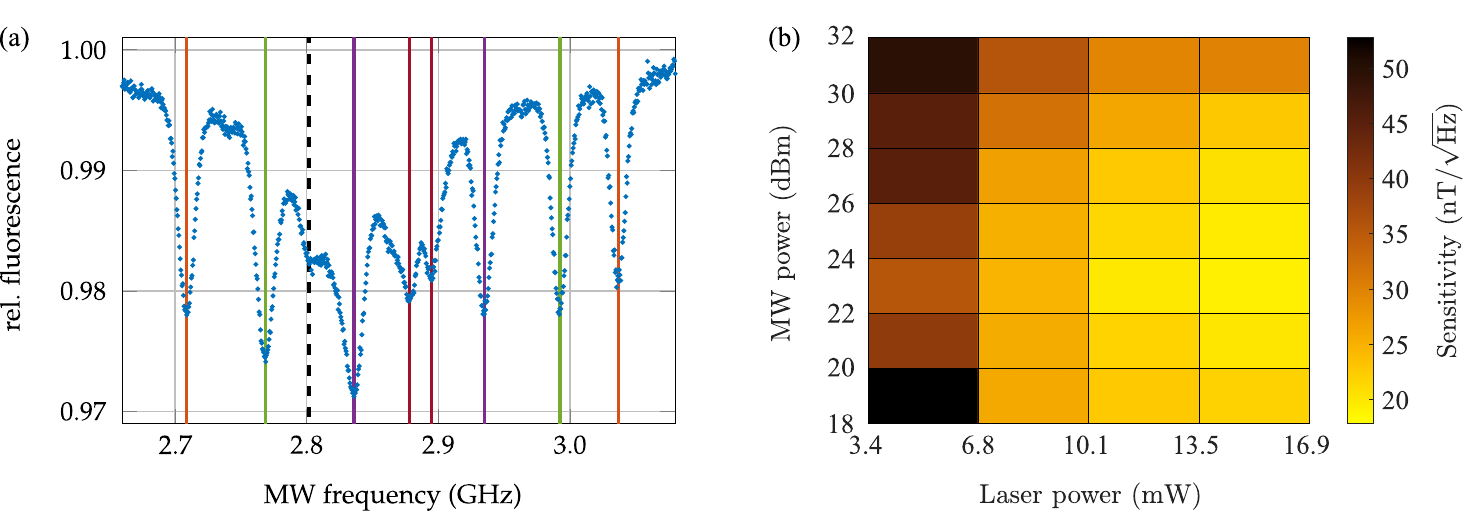}
		\caption{(a) ODMR spectrum of a single-crystalline microdiamond in an external magnetic field measured through the optical fiber. Eight resonances corresponding to the two spin transitions and four crystal axes are observed. Resonances of the same crystal axis are highlighted in the same color. Additionally, a ninth resonance at the fiber-tip antenna's resonance is observed in the spectrum (black dashed line). (b) Calculated sensitivity in dependence of Laser and MW power of the resonance at $\SI{2.71}{\giga\hertz}$ measured through the optical fiber.}
		\label{fig:ODMR_spectrum}
\end{figure*}

Using the endoscope fiber, we record the fluorescence intensity $I$ with and without applying the external microwave field through the endoscope antenna, and calculate the relative fluorescence level detected with the endoscope and the microscope simultaneously.
For each device the relative fluorescence level is calculated in dependence of the MW frequency $f_{\mathrm{MW}}$ as
\begin{equation}
    R(f_{\mathrm{MW}})= \frac{I_{\mathrm{MW on}}}{I_{\mathrm{MW off}}},
\end{equation}{}
where the index denotes if the state of the microwave source is turned on or off. 
The relative fluorescence level in dependence of the MW frequency yields an ODMR spectrum, which is fully excited and measured through the fiber. Such a spectrum is presented in FIG. \ref{fig:ODMR_spectrum} (a).

We observe eight resonances corresponding to four NV-axes with two spin-transitions $m_s = 0 \Longleftrightarrow m_s = -1$ and $m_s = 0 \Longleftrightarrow m_s = +1$ with both detection devices, i.e. the microscope and the fiber endoscope.
Further, we observe a feature in the fluorescence spectrum around $\SI{2.80}{\giga\hertz}$, which we attribute to the sharp resonance of the antenna (see section \ref{sec:characterization}) being in close proximity to the spin resonances and resulting in power-broadening.
The small shift of the antenna's resonance frequency with respect to the resonance frequency shown in FIG. \ref{fig:antennachar}(a) might stem from a different bending of the connection wires and the dielectric glass-substrate placed at small distance to the antenna.
So far, we have verified that ODMR can be achieved with the fabricated endoscope.
However, the benefit of the DLW structure connecting the wires for sensing applications has not been discussed. 
Therefore, we have compared the fabricated device to another endoscope, where the DLW process has been omitted, i.e., the two silver wires are not connected by a DLW structure, in appendix \ref{app:comparison_to_no_DLW}. 
We observe almost no ODMR signal from such a device when applying it to the same diamond under similar conditions.
This observation is in accordance with our expectation that the current within the metallic DLW structure mainly contributes to the magnetic field component at the diamond's position below the fiber core.
We deduce that a high coupling of the magnetic field component can thus be achieved with the presented device housing the DLW structure on the fiber-facet.

An important figure of merit to evaluate and compare the performance of the fiber endoscope is the magnetic field sensitivity $\eta$. 
For the individual resonances $i$ this quantity is defined as \cite{Dreau.2011}
\begin{equation}
    \eta_i = P_L \frac{1}{g_s \mu_B} \frac{w_i}{C_i \sqrt{R}},
\end{equation}
where $P_L=\frac{4}{3\sqrt{3}}$ is a factor for the maximum slope of a Lorentzian profile, $w_i$ is the width and $C_i$ is the relative fluorescence contrast of the respective resonance and $R$ is the photon-count rate.
Here, the width $w$ and relative fluorescence contrast $C$ depend on the driving-field strengths of the laser and microwave antenna, which can be optimized to achieve a minimum, i.e., an optimal sensitivity. 
The photon-count rate $R$ depends on the laser-intensity and total photon-collection efficiency.
Neglecting the effects of residual fiber fluorescence and experimental imperfections we estimate a ratio of total collection efficiency of 5:1 between confocal setup and fiber tip endoscope. 
After optimization of laser- and microwave-power to the resonance at $\SI{2.7145}{\giga\hertz}$, we achieve a sensitivity of $\eta_\mathrm{CM} = \SI{2.1}{\nano\tesla\per\sqrt{\hertz}}$ using the confocal microscope and $\eta_\mathrm{FT} = \SI{17.8}{\nano\tesla\per\sqrt{\hertz}}$ using the fiber-tip endoscope respectively.
This difference depends on the difference in the photon count-rate which is 70.3 times higher when using the confocal microscope. 
The difference of measured count rates is larger than the expected ratio of coupling efficiencies. This stems from imperfect fiber coupling between the endoscope fiber and the connected SPCM which we also attribute to the fact that the endoscope fiber and fiber coupler are not chromatically corrected.
In contrast, the fiber-coupling to the SPCM in the microscope setup was achieved by chromatic corrected microscope objectives.
A direct comparison between the fluorescence intensity coupled into the microscope objective and into the fiber endoscope shows a ratio of 6:1, measured directly in the beampath with a powermeter. This ratio is close to the expected value of 5:1 given by the different numerical apertures ($\mathrm{NA_{Microscopeobjective}}=0.50$; $\mathrm{NA_{Endoscope}}=0.22$). 
While the sensitivity measurement above was taken with the SPCM and additional ND filters ($\mathrm{OD_{Microscope}}=7.1$ and $\mathrm{OD_{Endoscope}}=5.6$), the fluorescence intensity coupled into the fiber-endoscope from a $\SI{15}{\micro\meter}$-sized diamond is high enough to be measured with an inexpensive photodiode as shown in appendix \ref{app:photodiode}.
In conclusion, compared to the microscope setup, our endoscope achieves sensitivities that differ by less than one order of magnitude due to the light coupling efficiency of both systems.

\subsection{Pulsed ODMR}

\begin{figure*}
    \centering
    \includegraphics[width=0.9\textwidth]{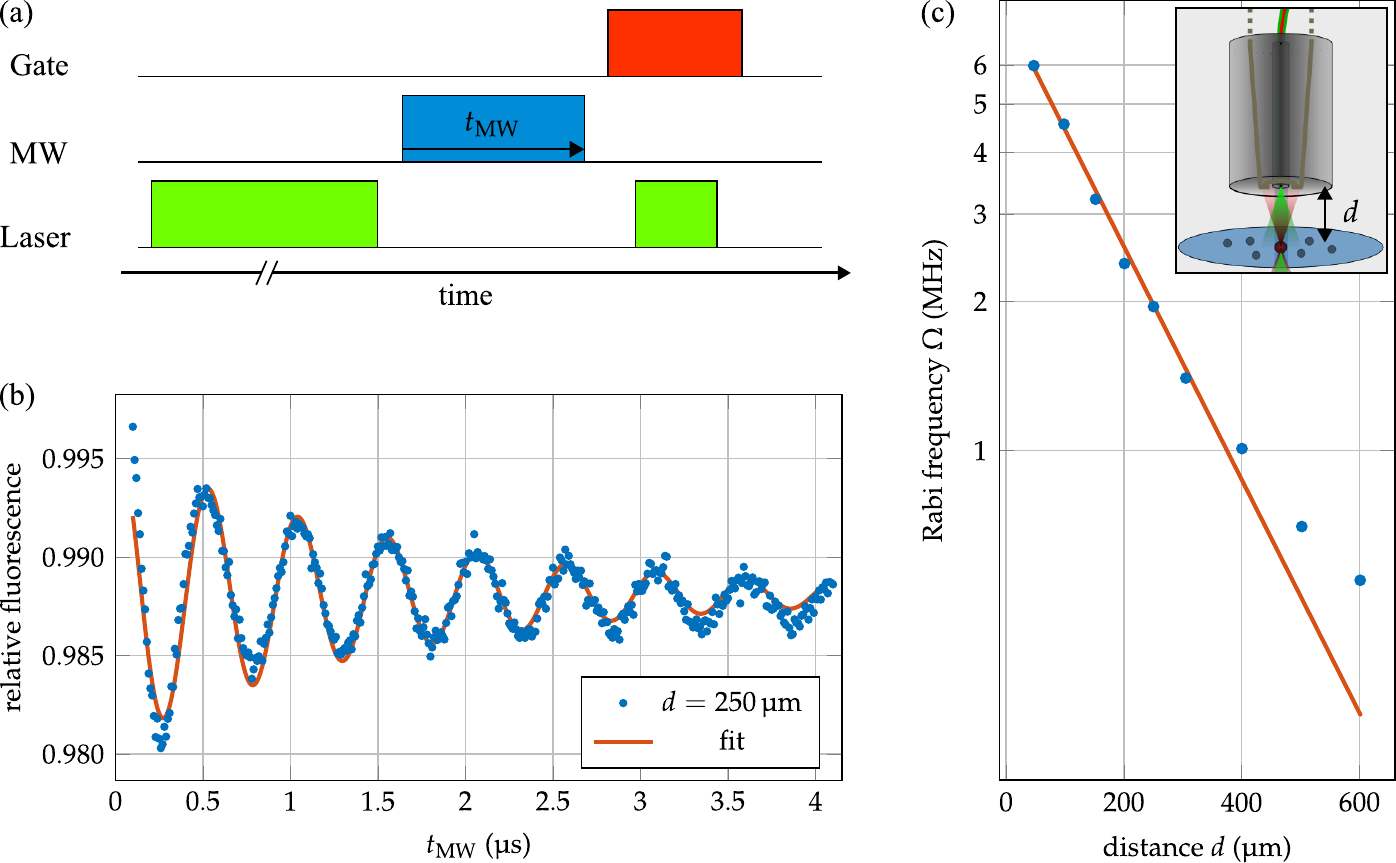}
    \caption{Rabi measurement in dependence of the antenna distance to the sample. (a) Pulse sequence of the Rabi cycle, (b) Rabi oscillations for a distance of $d=\SI{250}{\micro\meter}$ and fit curve to extract the Rabi frequency $\Omega$, (c) Extracted Rabi frequencies as a function of the distance $d$ and exponential fit; the inset shows a zoom-in to the measurement setup where the distance $d$ between the sample and the antenna is varied. The errors from the fit and the assumed measurement errors for the distance are smaller than the size of the data points.}
    \label{fig:ODMR_pulsed}
\end{figure*}

We use coherent Rabi oscillations in pulsed ODMR sequences to measure the coupling strength of the magnetic microwave field to the NV centers' spins in dependence of the distance of fiber tip and microdiamond.
The fiber tip endoscope is precisely positioned, i.e. the fiber core is centered in $x$ and $y$ direction above a microdiamond, with the wide-field microscope setup using an objective with $\SI{2.1}{\milli\meter}$ working distance and a depth of field of $\SI{4.2}{\micro\meter}$ at $\SI{650}{\nano\meter}$.
We control the distance between the endoscope's facet bearing the antenna and the microdiamond on a glass substrate using the microscope setup's motorized stage to infer the distance between sharp images of antenna and diamond.
At each distance we use pulsed ODMR to measure coherent Rabi oscillations of a single magnetic-field-splitted resonance with the sequence depicted in FIG. \ref{fig:ODMR_pulsed} (a).
The envelope of the rectangular microwave pulse is generated with digital In-phase and Quadrature-modulation (IQ) and the amplitude of the signal is set constant at $\SI{37}{dBm}$ directly at the output of the microwave amplifier.
Typical measurement data are shown in FIG. \ref{fig:ODMR_pulsed} (b). 
We infer the Rabi frequency $\Omega$, which is proportional to the field $B_\mathrm{MW}$ produced by the antenna at the microdiamond's position from a fit function according to
\begin{equation}
    f(t_{\mathrm{MW}})=y_0 + A \cos{(2 \pi \Omega t_{\mathrm{MW}} +\phi_0)} \exp{(-t_{\mathrm{MW}}/\tau)}
\end{equation}
of each data-set, where $y_0$ is the offset, $\Omega$ is the Rabi frequency, $\phi_0$ is a phase offset due to pulse imperfections and $\tau$ is the dephasing time of the spin ensemble during the Rabi cycle. The resulting Rabi frequency $\Omega$ in dependence of the distance is plotted in FIG. \ref{fig:ODMR_pulsed} (c). 

We observe a reduction of the Rabi frequency with increasing distance following phenomenologically an exponential law for small distances. 
The measured Rabi-frequency depends on the magnetic microwave field strength along the NV-axis and will therefore change for other orientations within the crystal.
We point out that the pulsed measurement scheme using Rabi oscillations could be applied to infer an unknown distance if the device is calibrated beforehand.

\section{Conclusion and Outlook}

In conclusion, we have presented the production of an electrical antenna on the facet of a multi-mode glass fiber using mDLW. 
The device combines electrical connections as well as optical access through the fiber on a diameter of the complete package of $d=\SI{2.5}{\milli\meter}$.
We have investigated the antenna characteristics by means of scattering parameters and found a high magnetic-field coupling in the near field. 
The device has been applied in a microscope setup for microdiamonds containing NV centers as an endoscope delivering optical and microwave excitation and guiding the fluorescence signal.

In comparison to a conventional microscope, the size of the integrated fiber-tip antenna is reduced by two orders of magnitude and can be further miniaturized.
The magnetic field sensitivity of $\SI{17.8}{\nano\tesla} / \sqrt{\SI{}{\hertz}}$ is decreased in comparison to a microscope setup by one order of magnitude due to the coupling, lower numerical aperture and losses in the fiber. Moreover an additional DLW lens structure on the fiber tip could improve the fluorescence collection and therefore increase the sensitivity.
In a complementary approach to mapping of antenna near fields on a sample plane with NV-centers \cite{Mariani.2020}, we have measured the reduction of the magnetic near field for varying distances in the micrometer regime.
Using a pulsed sequence for the optical detection of Rabi oscillations, the device allows the application as a distance sensor based on the magnetic field strength between the antenna and a quantum emitter showing spin-dependent fluorescence. 
We point out that the non-linear increase of the measured Rabi frequency with decreasing distance leads to higher precision for smaller distances.

Furthermore, the sharp resonances and small bandwidth achieved with the antenna can be favourable in systems, where a single frequency is required. 
Specifically, such a fiber-integrated antenna could extend the work on fiber-based cold atom studies \cite{Langbecker.2018,Vetsch.2010} by microwave near fields for hyperfine manipulation.
For magnetometry using NV-centers, however, a large bandwidth is favourable to reduce the applied current for all resonances. 
Further optimization of the mDLW structure, the galvanization and connection wires could allow tailoring the bandwidth for this purpose.

\section*{Acknowledgements}
We acknowledge support by the nano-structuring center NSC. This project is funded by the Deutsche Forschungsgemeinschaft (DFG, German Research Fundation) - Project-ID 454931666 as well as via the CRC 926 MICOS Project-ID 172116086 subproject B11. Furthermore, we thank Dennis Lönard and Isabel Manes for helpful discussions and experimental support.

\section*{Author declarations}
\subsection*{Conflict of Interest}
The authors have no conflicts to disclose.
\subsection*{Author Contributions}
The manuscript was written through equal contribution of S. Dix and J. Gutsche. All authors have given approval to the final version of the manuscript.

\section*{Data availability}
The data that support the findings of this study are available from the corresponding author upon reasonable request.

\appendix

\section{2-port measurements with two antennas in close proximity} \label{app:s_parameter}

\begin{figure*}
	    \centering
		\includegraphics[width=0.9\textwidth]{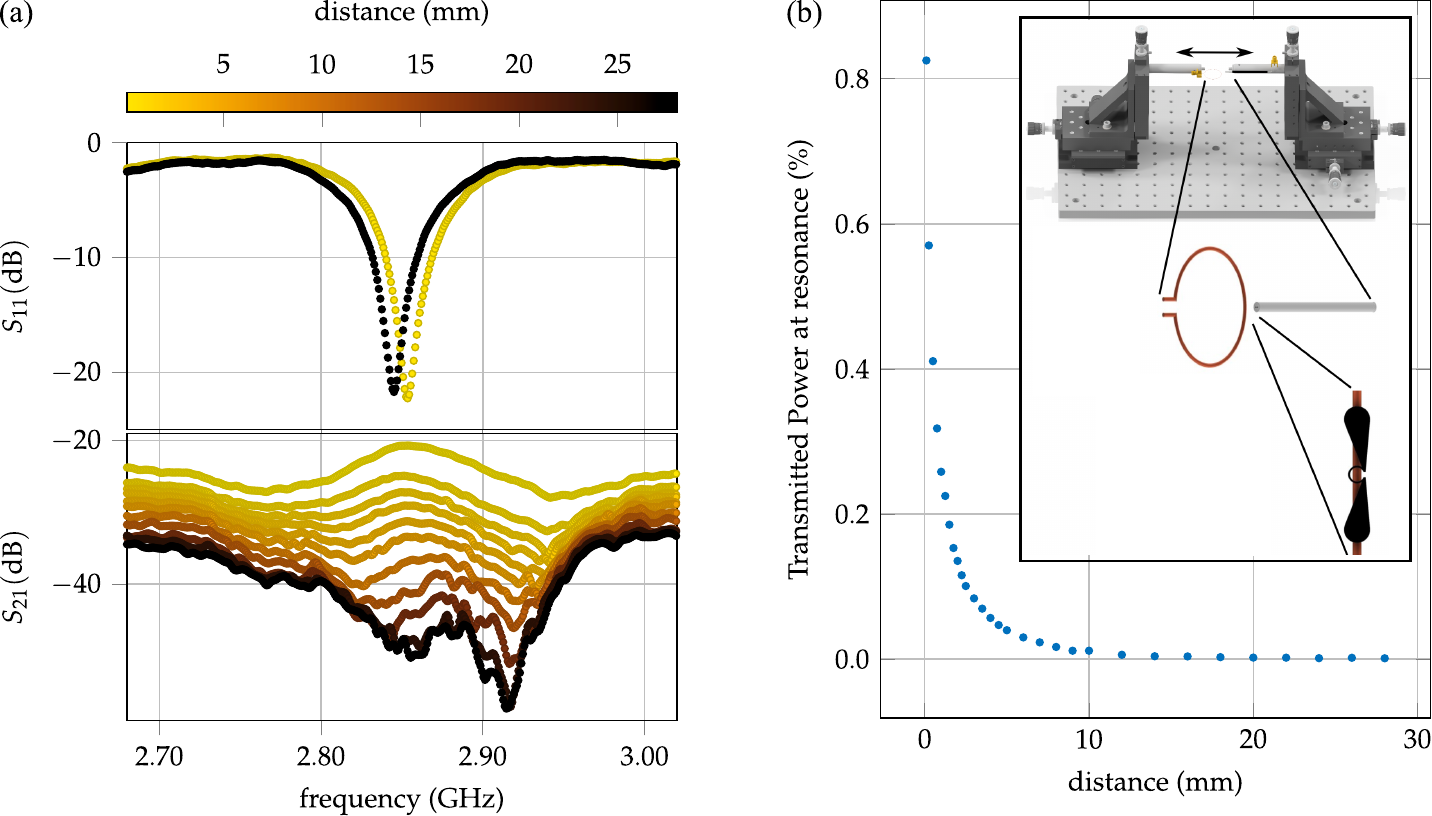}
		\caption{Experimental scattering parameters. (a) $S_{11}$ and $S_{21}$ parameter in dependence of the distance between the manufactured antenna on the fiber-tip to a reference $\lambda$ loop antenna. (b) Transmitted power at the resonance of the fiber-tip antenna in dependence of the distance. The inset shows the measurement setup and relative antenna alignment.}
		\label{fig:antennachar_app}
\end{figure*}
In addition to the $S_{11}$ parameter measurement, shown in FIG. \ref{fig:antennachar}, we also measure the scattering parameter $S_{11}$ and $S_{21}$ in a two-port measurement for varying distances between the fiber-tip antenna produced (1) and a reference loop antenna (2) with a circumference of $l = \lambda_\mathrm{MW} \approx \SI{10}{\centi\meter}$ and resonance frequency of $\nu_\mathrm{2,res}=\SI{2.87}{\giga\hertz}$, which is shown in FIG. \ref{fig:antennachar_app}. Specifically, $S_{21}$ represents the transmitted power from the antenna produced into the reference loop antenna. A schematic of the measurement setup for both antennas is presented in the inset of FIG. \ref{fig:antennachar_app} (b). The smallest distance between both antennas spans the positions of highest current at the resonance frequency in each antenna. 
Therefore, the coupling of the magnetic near field is expected to be the dominant contribution to the measured transmission, i.e. the $S_\mathrm{21}$ parameter.
The distance between both antennas was set to zero by bringing both antennas to a visible contact where only the coating of the second antenna around the copper wire separates both antennas. 
We measure the $S_{21}$-parameter for varying distances between both antennas to track the distance dependence of the transmitted power in the near field. The measurement data is shown in FIG. \ref{fig:antennachar} (a), the data points at the resonance frequency $\nu_\mathrm{1,res}=\SI{2.846}{\giga\hertz}$ are plotted versus the distance in FIG. \ref{fig:antennachar} (b).
We observe a rapid reduction of the transmitted power with increasing distance at small distances. 
We therefore conclude that the produced fiber-tip antenna transmits a relatively high amount of the power in close proximity to the antenna.
However, at such small distances both antennas can not be assumed to be point-like magnetic dipoles. 
Hence, this measurement is prone to errors in comparison to the proposed application, where only the magnetic field component emitted by the antenna drives a spin transition.

\section{Comparison to a device without DLW structure}
\label{app:comparison_to_no_DLW}

\begin{figure}
	    \centering
		\includegraphics[width=0.45\textwidth]{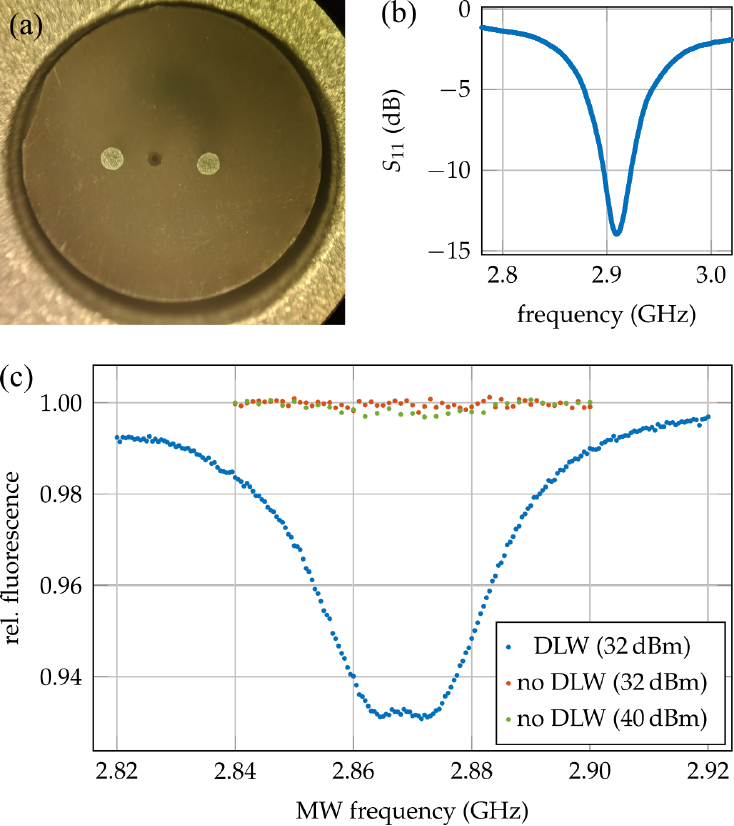}
		\caption{\textbf{(a)} Facet of a fiber-tip endoscope without DLW structure. \textbf{(b)} Scattering parameter $S_{11}$ of such an endoscope measured with a VNA. \textbf{(c)} Comparison of ODMR-spectra taken with the endoscope featuring the DLW connection (blue) and without such a structure at different microwave power (orange, green).}
		\label{fig:unconnected_antenna}
\end{figure}

In order to analyze the benefit of the DLW sturcture connecting the silver wires of the fiber-tip endoscope, we compare it to a device lacking the DLW step in the production. 
In FIG. \ref{fig:unconnected_antenna} (a), we show the facet of such an endoscope.

We have measured the reflection coefficient $S_{11}$ of the fiber without the metallic DLW connection and show the measurement data in FIG. \ref{fig:unconnected_antenna} (b).
The resonance frequency is $\nu_{\mathrm{res}}=\SI{2.909}{\giga\hertz}$ and the minimum value for the reflection coefficient is $S_{11}(f)=\SI{-14}{dB}$. The bandwidth amounts to $\Delta \nu_{\mathrm{res, \SI{-3}{dB}}} = \SI{21.8}{\mega\hertz}$.
In comparison to the endoscope featuring the DLW connection (see section \ref{sec:characterization}), the resonance frequency is also in close proximity to the NV centers' resonance.
Moreover, the scattering parameter $S_{11}$ does not show such a strong reduction and a slightly larger $\SI{3}{dB}$-width.
We conclude that the electrical resonance of the antenna is mainly given by the silver wires.

Furthermore, we have applied both endoscopes as an antenna for measuring an ODMR spectrum in the absence of a magnetic field. In this case, we used the microscope setup for optical excitation and detection. 
We only applied the endoscope as an antenna to rule out an impact on the results due to other differences in excitation or detection efficiency.
We used the same laser power, MW amplitude and distance between fiber facet and diamond for this comparison.
The resulting ODMR spectra are shown in FIG. \ref{fig:unconnected_antenna} (c).
We observe the expected two resonances in the absence of a static magnetic field for the endoscope with DLW structure.
For the endoscope without such a structure, we observe an almost flat line and no reduction of fluorescence within the noise of the measurement. 
We have further increased the MW amplitude to $\SI{40}{dBm}$ (by a factor of $\SI{6.3}{}$) and repeated the measurement. 
In this case a very small reduction of the intensity can be observed, which is still much less than observed with the endoscope housing the DLW structure.
We therefore conclude that the current in the DLW structure is mainly contributing to the microwave's magnetic field component for such small distances. 
This enables an efficient microwave excitation of spin states due to the metallic DLW structure reducing the applied microwave power needed to drive the spin states. 

\section{ODMR measurement with a photodiode \label{app:photodiode}}
\begin{figure}[htbp]
	    \centering
		\vspace{0.02\textwidth}
	    \includegraphics[width=0.45\textwidth]{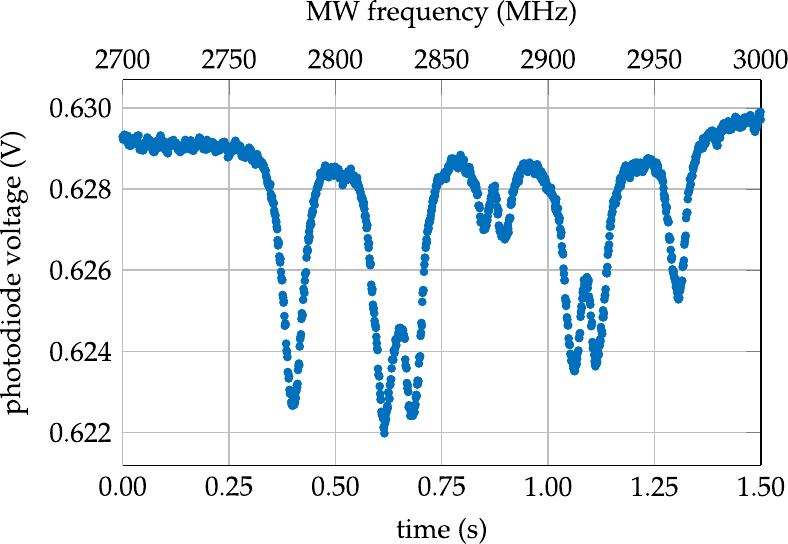}
		\caption{ODMR spectrum taken with a photodiode (DET36A/M from Thorlabs). The parameters for the sweep of the microwave were step time $\mathrm{t_{step}}=\SI{4}{\milli\second}$ and step width $\mathrm{w_{freq}}=\SI{0.8}{\mega\hertz}$.}
		\label{ODMR PD}
\end{figure}

Instead of using a SPCM with several grey filters to avoid saturation of the device we also applied a photodiode to detect the NV centers' fluorescence of a single microdiamond. The excitation laser light and MW signal is supplied continuously and the MW frequency is ramped linearly from $\SI{2.7}{\giga\hertz}$ to $\SI{3.0}{\giga\hertz}$ in $\SI{1.5}{\second}$ and is limited by the minimal time step of the microwave generator (SynthHD (v2) from Windfreak Technologies). The photodiode signal (voltage) measuring the spin dependent fluorescence is monitored on an oscilloscope. The measured ODMR signal is shown in FIG. \ref{ODMR PD}. We observe the typical 8 resonances produced by a single crystalline diamond. To correlate the time dependent x-axis to a frequency dependent axis, the distances between the corresponding resonances $m_s=0$ to $m_s=+1$ and $m_s=0$ to $m_s=-1$ are fitted for each NV-axis as a time difference. With the chosen step time and step width we convert the time distance between resonances to frequency differences leading to a frequency dependent x-axis with an unknown offset. By modeling all possible resonance frequencies for all NV-axis orientations and external magnetic field orientations and subsequently comparing the distance between the resonance positions (measured and modelled), we are able to find fixed points for the resonances. These are then used to calculate the offset of the frequency dependent x-axis.
\bibliography{fiber_bib}

\end{document}